# BPM in the cloud: A systematic literature review

A new approach on BPM in the cloud. Elastic Business Process Management as a Service (eBPMaaS)

Ageo Carrillo
Systems and Informatics Engineering Faculty
UNMSM
Lima, Perú
ageo.carrillo@unmsm.edu.pe

Marco Sobrevilla
Systems and Informatics Engineering Faculty
UNMSM
Lima, Perú
msobrevillac@gmail.com

*Abstract*—Business Process Management (BPM) in the cloud is focused on how to provide Business Process as a Service (BPaaS) and also implementing the Elastic Business Process Management (eBPM). These approaches are finding models, techniques and methodologies to increase the BPM adoption in the cloud. The objective on this research is to carry out a literature review about the state of the art of Business Process Management in the cloud, identify categories and analyze what the approach on eBPM researches and BPaaS researches is. The method used is the systematic literature review and the researches evolution over the time will be categorized and analyzed using a timeline and cumulative charts. The results show that the Business Process enactment is the main approach focused on eBPM researches, whereas the researches on BPaaS are focused on a greater number of categories. The conclusion of this research is that the authors could not identify any systems supporting the eBPM and BPaaS approaches at the same time and they introduce the definition of a new approach called Elastic Business Process Management as a Service (eBPMaaS)

## I. INTRODUCTION

Business Process Management (BPM) is a discipline adopted by organizations to improve their productivity by integrating people, systems, information and things in order to achieve the strategic objectives of the organization [41]. [1] presents the "Elastic Processes" concept and defines it as the processes in which machines and people work together to perform data processing tasks complementing each other to obtain results using elastic computing units. Based on the Elastic Processes definition, [3] presents the architecture to implement Elastic Business Process Management Systems (eBPMSs) that has an intelligent component which allows rationalizing the work on elastic units based on quality of services parameters. On the other hand, cloud computing is a type of technology that allows organizations to consume infrastructure, platform and software as a service under the pay-as-you-go model. In this context, in [4] they present a classification of services in the cloud with an additional model called Business Process as a Service (BPaaS) that provides automated processes in the cloud and that can be consumed as a service under the pay-as-you-go model.

The main approach on researches related to Elastic Business Process Management and Business Process as a Service is increasing the adoption of business processes in the cloud. Each approach addresses different important issues that facilitates the adoption of business processes in the cloud and serves as a tool for Business Process Management in the cloud.

## I. BACKGROUND AND RELATED WORK

In [1] the principles of the elastic processes are brought up from the premise that machines can process high data volumes and obtain results in short time, but in some cases the involvement of the human being in the data analysis is necessary in order to obtain conclusions because that can't be determined by the machines. With this principle, the difference between machine-computing units and human-computing units is made and they must work together to process data and obtain results. For them to work together it is necessary to establish a type of process which counts with an intelligent component that controls the work between units of machine computing and human computing taking into account the use of resources and quality and cost requirements. These type processes are defined as Elastic Processes.

In [2] elasticity is defined as a computing feature that refers to the property a system has in order to increase and decrease its processing capacity. The elasticity in computing is performed through the virtualization of computation units. The virtualization in machines is used through cloud computing while virtualization in humans is provided by crowdsourcing. In [2] an example of image patterns discovering process is proposed in which machines can identify patterns, however in some cases it is necessary the intervention of a human to detect the correct pattern or to make a validation regarding the pattern detected by the machine.

In [3] the definition of systems that allow to manage elastic processes is presented and named Elastic Business Process Management Systems (eBPMSs). From this definition a high-level architecture to these type of systems is presented and an eBPM meta-model is proposed. The high-level architecture considers the use of virtual machines for interaction between computing units, in the case of machine computing units, it provides web services and for human computing it provides web applications, each of these interfaces communicates with a virtual machine that manages automatic tasks for machine computing units (MCU) and human tasks for human computing units (HCU). The tasks are executed at the request of a system that manages its execution based on the parameters of quality and service defined for the process.

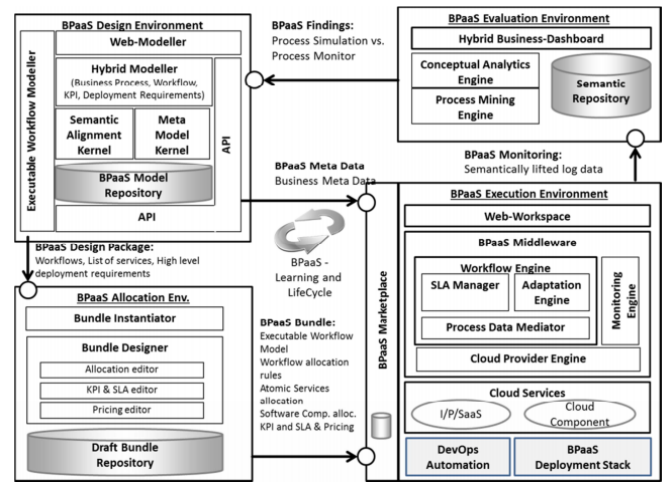

Fig. 2. High level Architecture of BPaaS Enviroments [3]

This high-level architecture serves as a basis for the implementation of systems that provide BPaaS and suggests the meta-model to implement a system called CloudSocket which is developed over the H2020 project of the European Union framework.

## II. RESEARCH METHOD

Based on the guidelines for performing Systematic Literature Review in software engineering [7] the question that arises is "What is the research approach on business processes in the cloud?". During the initial literature review, two main terms were identified to refer to business processes in the cloud: elastic Business Process Management (eBPM) and Business Process as a Service (BPaaS); both focused on overcoming the challenges implementing business processes in the cloud.

The research process consists of three phases:

- Article Selection.
- Development of the systematic map
- Results

### A. Article Selection

Based on the guidelines for performing Systematic Literature Review in software engineering the authors searched for information in the following academic article search engines: Science Direct and Google Scholar. The search was conducted based on two approaches, the first related to eBPM and the second related to BPaaS.

For the search focused on eBPM, the following query was used:

- "Elastic Business Process"

37 results were generated from the search, 20 of these were left out because there were duplicates or some of them

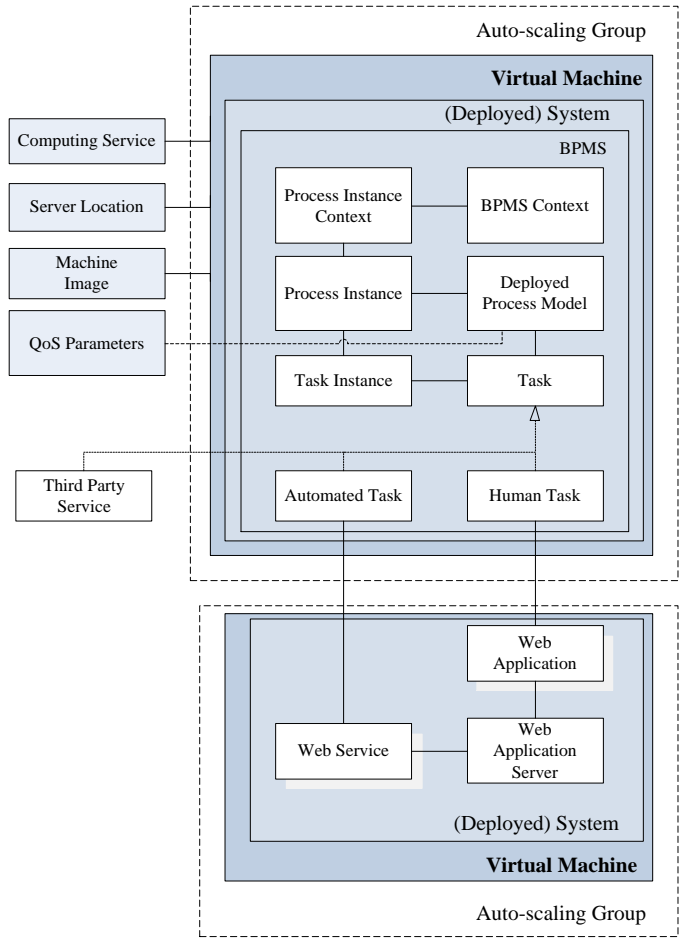

Fig. 1. High level architecture eBPMSs. Source: Adapted from [3]

On the other hand, cloud computing is a technology with increasing adoption due to the possibility of consume computer services under the pay-as-you-go model. According to NIST [40] there are three models to consume computation in the cloud, these are Infrastructure as a Service (IaaS), Platforms as a Service (PaaS) and Software as a Service (SaaS). In [4] they refer to these models and present in their classification an additional service model called Business Process as a Service (BPaaS) that provides business processes that connect to different systems created on the SaaS model. According to [5] this allows companies to be more agile in the sense that they can have a well-defined, granular and consistent piece of technology defined according to the policies and processes of the companies. On this concept, [6] presents the high-level architecture for BPaaS systems, which is composed of four environments: design, assignment, execution and evaluation.

did not present concrete results; after the selection, 17 articles are presented in TABLE I.

TABLE I. eBPM RESEARCH RESULTS

| Ref | Research | |
|---|---|---|
| | Title | Year |
| [8] | The PADRES Distributed Publish / Subscribe System | 2005 |
| [9] | A Distributed Service-Oriented Architecture for Business Process Execution | 2010 |
| [1] | Principles of elastic processes | 2011 |
| [2] | Virtualizing Software and Humans for Elastic Processes in Multiple Clouds | 2012 |
| [10] | Self-adaptive resource allocation for elastic process execution | 2013 |
| [11] | Workflow scheduling and resource allocation for cloud-based execution of elastic processes | 2013 |
| [12] | Realizing elastic processes with viePEP | 2013 |
| [13] | Introducing the vienna platform for elastic processes | 2013 |
| [14] | Cost-driven Optimization of Cloud Resource Allocation for Elastic Processes | 2013 |
| [15] | ViePEP - A BPMS for elastic processes | 2014 |
| [16] | Elastic Multi-tenant Business Process Based Service Pattern in Cloud Computing | 2014 |
| [17] | Cost-Efficient Scheduling of Elastic Processes in Hybrid Clouds | 2015 |
| [18] | Modeling, Evaluation and Provisioning of Elastic Service-based Business Processes in the Cloud | 2015 |
| [19] | Optimization of Complex Elastic Processes | 2015 |
| [20] | Four-Fold Auto-Scaling on a Contemporary Deployment Platform Using Docker Containers | 2015 |
| [3] | Elastic business process management: state of the art and open challenges for BPM in the cloud | 2015 |
| [21] | An Efficient Approach for Multi-tenant Elastic Business Processes Management in Cloud Computing environment | 2016 |

For the search focused on BPaaS, the following query was used:

- Allintitle: Cloud Business Process
- Allintitle: Business Process in the cloud
- Allintitle: Business Process as a Service
- Allintitle: BPaaS

The search generated 39 results, 18 of these were excluded because in some cases the article was focused on the business process, not the technology that supports the process and in other cases it was focused on a specific product without focusing on the outcome of the research; after this selection, 21 articles are presented on TABLE II.

TABLE II. BPaaS RESEARCH RESULTS

| Ref | Research | |
|---|---|---|
| | Title | Year |
| [22] | Leveraging business process as a service with blueprinting | 2011 |
| [23] | Integrating Workflow-Based Mobile Agents with Cloud Business Process Management Systems | 2012 |
| [24] | A cloud HUB for brokering business processes as a service: A 'rendezvous' platform that supports semi-automated background checked partner discovery for cross-enterprise collaboration | 2012 |
| [25] | Business Process Management in the cloud : Business Process as a Service (BPaaS) | 2012 |
| [26] | Anonyfrag : An Anonymization-Based Approach For Privacy-Preserving BPaaS Categories and Subject Descriptors | 2012 |
| [27] | A Modelling Environment for Business Process as a Service | 2013 |
| [28] | Building a customizable business-process-as-a-service application with current state-of-practice | 2013 |
| [29] | Business Process as a Service: Chances for Remote Auditing | 2013 |
| [4] | Business Process as a Service – Status and Architecture | 2013 |
| [30] | A Security Risk Assessment Model for Business Process Deployment in the Cloud | 2014 |
| [31] | Separating Execution and Data Management: A Key to Business-Process-as-a-Service (BPaaS) | 2014 |
| [5] | Towards a Framework for Defining and Categorizing Business Process-as-a-Service (BPaaS) | 2014 |
| [32] | jBPM4S: A Multi-tenant Extension of jBPM to Support BPaaS | 2014 |
| [33] | A New Framework for Cloud Business Process Management | 2014 |
| [34] | A Context Based Scheduling Approach for Adaptive Business Process in the Cloud | 2014 |
| [35] | Managing Configurable Business Process as a Service to Satisfy Client Transactional Requirements | 2015 |
| [36] | A study on BPaaS with TCO model | 2015 |
| [6] | Business Process as a Service (BPaaS) The BPaaS Design Environment | 2015 |
| [37] | A semantic framework for configurable business process as a service in the cloud | 2016 |
| [38] | BPaaS Modelling : Business and IT - Cloud Alignment based on ADOxx | 2016 |
| [39] | Security-aware Business Process as a Service by hiding provenance | 2016 |

*B. Development of the systematic map*

To identify differentiated approaches among the selected articles, the articles were grouped into seven categories: Design, Enactment, Evaluation, Costs, Security, Terminology and Adoption.

*1) Design:* Researches focused on the study of frameworks, models and techniques for the business processes design.

*2) Enactment:* Researches focused on optimizing hardware resources consumption when running business process management systems in the cloud. They present mathematical models for hardware cost calculation and reduction in hardware consumption in the cloud and also algorithms of cost minimization that can be implemented in hardware resources reasoning systems in the cloud.

*3) Evaluation:* Researches focused on giving solutions that allow the business processes evaluation when its implemented in the cloud. It analyzes the opportunities to perform a remote audit to the business processes when being implemented in the cloud and the separation of execution and data management is presented to facilitate the work of auditing business processes in the cloud.

*4) Cost:* Researches focused on the costs in business processes execution in the cloud. These researches present a mathematical model for the Total Cost Ownership (TCO) evaluation in Business Processes as a Service and apply the model to a real case. As a result, the possibility to determine the Total Cost of Ownership is validated.

*5) Security:* Researches focused on data security that are part of business process management when implemented in the cloud and security delivered by cloud providers. Solutions are designed to ensure data confidentiality by implementing business processes in the cloud through techniques to hide data through anonymous views and create diverse views of the business process.

*6) Terminology:* Researches on this category propose and define new concepts and terms used when referring to the management of elastic business processes and business processes as a service categorization. The defined terms serve to maintain a controlled language on the business process management in the cloud domain.

*7) Adoption:* The research in this category performs a field-based analysis to determine the adoption level of business processes in the cloud and the benefits it offers to the end users.

To compare work related to Elastic Business Process Management, we group the researches by classification category and publication year in Table III.

TABLE III. CATEGORIZED EBPM RESEARCHS RESULTS

| Category | Research | |
|---|---|---|
| | *References* | *Year* |
| Enactment | [8] | 2005 |
| | [9] | 2010 |
| | [10][11][12][13][14] | 2013 |
| | [15][16] | 2014 |
| | [17][18][19][20] | 2015 |
| | [21] | 2016 |
| Terminology | [1] | 2011 |

| Category | Research | |
|---|---|---|
| | *References* | *Year* |
| | [2] | 2012 |
| | [3] | 2015 |

When analyzing the researches related to Elastic Business Process Management by categories, it is possible to identify that these works are focused on providing support to the enactment phase in the Business Process Management lifecycle. These works are mainly focused on the changes which should be applied to the software architecture of the Business Process Management Systems for their execution in the cloud and take advantage of the cloud elasticity to reduce the costs in the hardware resources consumption; this can be done through mathematical models to determine the hardware resources consumption and to reduce consumption through minimization algorithms.

The terminology category presents terms for the definition of business processes in the cloud as "elastic processes", "elastic business process management" and "elastic business process management systems". These works allow the establishment of a controlled language to refer to the management of elastic business processes and in general to refer to the business processes in the cloud; these defined terms can be the basis for the elaboration of an ontology in the business processes in the cloud domain.

In order to compare related work to Business Process as a Service, the researches were grouped by classification category and publication year in Table IV.

TABLE IV. CATEGORIZED BPAAS RESEARCHS RESULTS

| Category | Research | |
|---|---|---|
| | *References* | *Year* |
| Design | [22] | 2011 |
| | [27] | 2013 |
| | [35] | 2015 |
| | [37][38] | 2016 |
| Enactment | [23][24] | 2012 |
| | [28][4] | 2013 |
| | [32][34] | 2014 |
| | [6] | 2015 |
| Evaluation | [29] | 2013 |
| | [31] | 2014 |
| Cost | [36] | 2015 |
| Security | [26] | 2012 |

| Category | Research | |
|---|---|---|
| | References | Year |
| | [30] | 2014 |
| | [39] | 2016 |
| Terminology | [25] | 2012 |
| | [5] | 2014 |
| Adoption | [33] | 2014 |

When analyzing the researches related to Business Processes as a Service by categories, it is possible to identify that these works are focused on a greater number of categories compared to the Elastic Business Processes Management; by covering more categories it was possible to cover more aspects related to business processes in the cloud.

Researches related to the design category present solutions to the challenge of including new functionalities that must be reflected in the business processes design due to their implementation in the cloud. For instance, configurable processes implementation to support different businesses processes, services description to which business process integrate in the cloud, mechanisms to design Business Processes as a Service and techniques for process design based on levels. The works in the enactment category are focused on the study of systems architecture designs that allow the implementation of business process management systems in the cloud and present the implementations results. The works in the evaluation category are focused on the opportunities offered by the business processes as a service implementation for the monitoring of business activities.

The costs category research presents a model for the calculation of the cost of ownership of the Business Processes as a Service and it applies the model to a real case to validate that with this model it is possible to calculate the total cost of ownership. Works in the security category focus on aspects of data security used during the execution of business processes in the cloud and present solutions to improve levels of accessibility to this data. The terminology category research presents terms for the definition of Business Processes in the cloud as "Business Process as a Service" and a categorization for Business Process as a Service over a framework called SEFIS. These researches allow the establishment of a controlled language to refer to the business processes in the cloud and are a source to establish the basis on the elaboration of an ontology of the domain of the business processes in the cloud. The research on the adoption category performs a quantitative and qualitative analysis to determine the level of adoption of business processes in the cloud and the benefits it offers to end users; this research is important as part of the literature to have an idea about the level of adoption and as a basis for future studies in other geographical regions.

*C. Results*

To make a timeline analysis based on the results of the literature review on Elastic Business Processes Management cumulative analysis charts are presented with the number of researches and categories identified by year. Figure 2 shows the eBPM timeline researches.

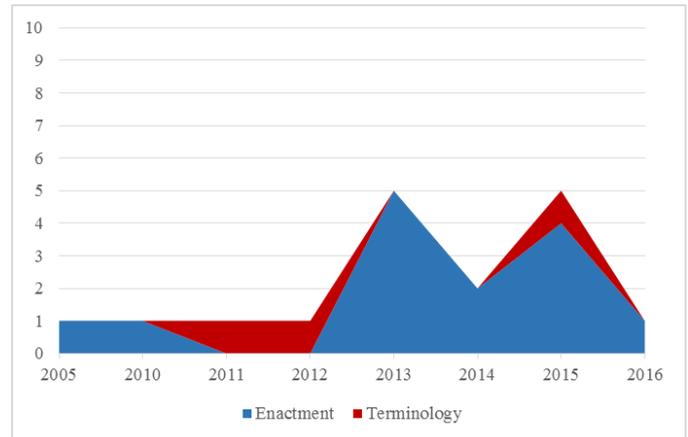

Fig. 3. Timeline eBPM researches.

From Figure 3 can be seen that the work related to Elastic Business Process Management has increased and they are mainly focused on the enactment of business processes in the cloud accompanied by some research in terminology that formulates new terms to refer to Elastic Business Process Management. It can be seen that they have not focused on any of the other categories, therefore, this is a short-range and highly specialized approach.

To make a timeline analysis based on the results in literature review on Business Process as a Service cumulative analysis charts are presented with the number of researches and categories identified by year. Figure 4 shows the BPaaS timeline researches

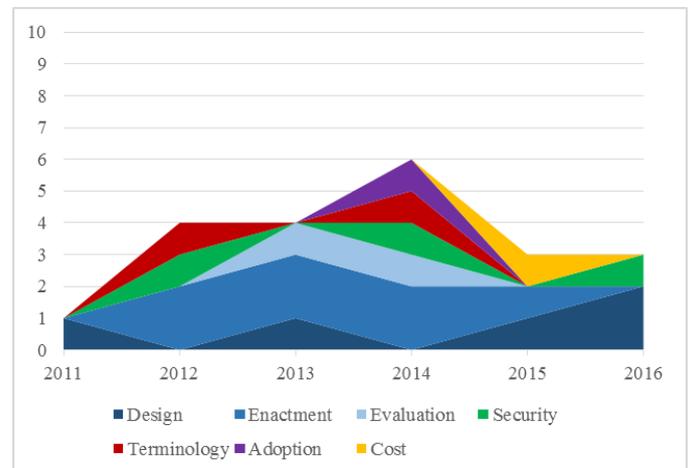

Fig. 4. Timeline BPaaS researches.

From Figure 4 we can see that the work related to Business Process as a Service has increased during 2014 and in the last years are mainly focused on improving the adoption levels in Business Process Management Systems in the cloud when proposing improvements that facilitate designing business processes in the cloud and increasing data security.

## II. CONCLUSION

The articles related to Elastic Business Process Management and Business Processes as a Service are carried out in each of the themes has focused more strongly on one of the seven categories. On Elastic Business Process Management researches the focus is enactment to overcome the infrastructure challenges when develop this systems in the cloud. On the other hand, the approach in Business Process Management as a Service researches covers a greater number of categories with a greater focus on improving the design and security of business processes in the cloud.

Figure 5 shows the timeline researches on BPaaS and eBPM, the same categories used in Figure 3 and Figure 4 have been applied and both approach researches have been consolidated to have an integral view of the researches done on business processes in the cloud.

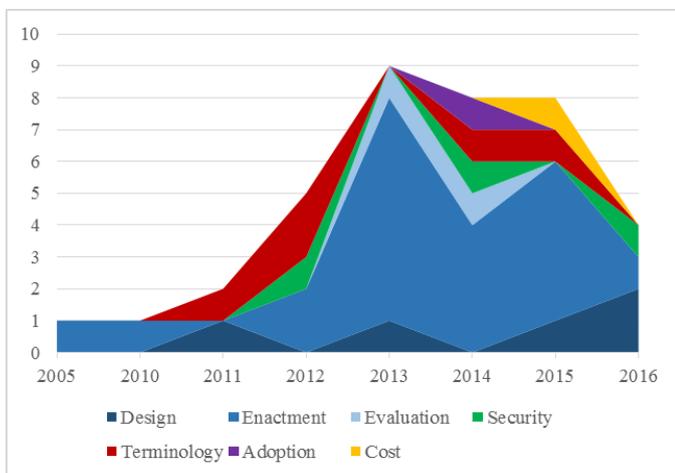

Fig. 5. Timeline eBPM and BPaaS researches.

Based on a literature review related to business processes in the cloud, it is concluded that Elastic Business Process Management approaches and Business Process as a Service continuously seek to improve the adoption of business processes in the cloud either to improve the management of business processes or to improve the systems architecture to reduce costs.

In the literature review, it was not possible to identify Elastic Business Process Management System that could allow the provision of Business Processes as a Service at the same time. The authors of this paper propose a new approach referred to Elastic Business Process Management Systems that at same time supports the provision of Business Processes as a Service, the definition to this new approach is Elastic Business Process Management as a Service (eBPMaaS). With this new approach, it will be possible to have a system ready to adopt the research results carried out on Elastic Business Process Management, which are aimed at making efficient use of hardware resources in the cloud focused on processor consumption, memory consumption and data transfer based on autonomic computing, auto-scaling and mathematical models on cost minimization. At the same time, the eBPMaaS approach will be prepared to adopt the results on investigations related to Business Process as a Service, which are focused on supporting the life cycle of business processes and functionalities that improve the security of the process data, develop configurable processes, remote audit, reduce total cost of ownership and support new process modeling techniques.

The future research directions on eBPMaaS define an eBPMaaS meta-model and a high-level architecture to eBPMaaS Systems. From the enactment approach, it is necessary to test the impact on elasticity when implementing the eBPMaaS System compared to traditional BPaaS Systems. From design approach, is necessary to establish a modeling language to support the eBPM and BPaaS approaches integrated, with a main approach on the design and enactment phases to implement new eBPMaaS systems.